# On-a-chip biosensing based on all-dielectric nanoresonators

*Ozlem Yavas[1], Mikael Svedendahl[1], Paulina Dobosz[1], Vanesa Sanz[1], and Romain Quidant[1,2]*

[1] ICFO-Institut de Ciencies Fotoniques, Barcelona Institute of Science and Technology, 08860 Castelldefels (Barcelona), Spain

[2] ICREA–Institució Catalana de Recerca i Estudis Avançats, 08010 Barcelona, Spain

Nanophotonics has become a key enabling technology in biomedicine with great promises in early diagnosis and less invasive therapies. In this context, the unique capability of plasmonic noble metal nanoparticles to concentrate light on the nanometer scale has widely contributed to biosensing and enhanced spectroscopy. Recently, high-refractive index dielectric nanostructures featuring low loss resonances have been proposed as a promising alternative to nanoplasmonics, potentially offering better sensing performances along with full compatibility with the microelectronics industry. In this letter we report the first demonstration of biosensing with silicon nanoresonators integrated in state-of-the-art microfluidics. Our lab-on-a-chip platform enables detecting Prostate Specific Antigen (PSA) cancer marker in human serum with a sensitivity that meets clinical needs. These performances are directly compared with its plasmonic counterpart based on gold nanorods. Our work opens new opportunities in the development of future point-of-care devices towards a more personalized healthcare.

The need for point-of-care devices in health and wellness monitoring is one of the principal motivations behind the current development in biosensing. Among the different available transduction schemes, optical biosensors hold the advantage of being highly sensitive, enabling label free and cost effective real time sensing[1]. Beyond silicon-based integrated optics[2–4] that shows great promises for sensing, surface plasmon resonance (SPR)[5–8] and fiber optics[9,10] based sensors utilizing propagating evanescent waves have been extensively studied and validated on a wide range of analytes. However, coupling incoming light to the surface modes usually requires sophisticated optics and such sensors are often limited to large bioanalytes, owing to a substantial spatial mismatch of the sensing mode with the tiniest molecules.

These limitations can partially be overcome by using 3D nano-optical resonators. In particular, extensive efforts have been put on localized surface plasmon resonance (LSPR) sensors[11–13] which exploit the unique properties of noble metal nanoantennas. The ability to excite LSPR with freely propagating incident light considerably simplifies the optical setup needed for such label free measurements. Highly confined modes also provide strong overlap between the electromagnetic field on the surface and the relevant biological analyte dimensions. Finally, the tiny size of each nanosensors enable their assembly in small foot print sensing areas compatible with parallel multi-detection platforms[11–13]. However, plasmonic nanoparticles suffer from dissipative losses in the metal that lead to broad resonances that eventually limit the sensitivity of the sensor read-out. Recently, high refractive index dielectric nanoparticles have been proposed as an attractive alternative to plasmonic nanoparticles in wide range of applications.[14,15]

All-dielectric nanophotonics is a fast progressing field which enables manipulation of both electric and magnetic components of the incoming light. These unique properties open up new opportunities in the field of metamaterials including negative refractive index, cloaking, superlensing and many more.[15–20] In practice, light coupling to dielectric subwavelength particles results in the excitation of both magnetic and electric multipole resonances which translates into multiple peaks in their extinction spectrum. Similar to metallic nanoparticles, the resonance frequencies depend on their geometry, constitutive material and the dielectric environment. Their sensitivity to the surrounding dielectric permittivity along with their strong mode localization suggests high index dielectric nanoparticles could perform well as biosensing transducers.[14,21–23] Silicon nanoresonators, with resonances in visible-NIR spectral range, were first studied theoretically and more recently measured experimentally[24–28]. The use of silicon is motivated by its compatibility with the microelectronics industry, high material abundance and low cost. While it has recently been suggested that Si nanoresonators could benefit to the detection of biological molecules, so far, only bulk refractive index sensing measurements have been reported.[21–23]

In this letter, we demonstrate that Si nanoresonator arrays integrated with state-of-the-art microfluidics result in an efficient sensing platform for the detection of protein cancer markers in human serum, at clinically relevant concentrations for cancer screening. We first study the optimal structural design of Si nanodisks for molecular sensing. Then, we demonstrate detection of PSA (prostate specific antigen) in buffer with a limit of detection (LOD) that is comparable to gold standard immunoassay techniques. Finally, to validate its operation in clinical conditions, the platform is tested in human serum.

Our platform consists of silicon nanodisk arrays on a quartz substrate integrated with a PDMS microfluidic chip including micromechanical valves (Fig. 1a-c). We fabricate the silicon nanodisk arrays using standard negative resist e-beam lithography followed by a reactive ion etching step on commercial silicon coated quartz samples (see the Supplementary Information). The nanodisk arrays have a fixed height $h$=50 nm. We choose to tune the disk radius $r$ and inter-particle distance $s$ to assess the optimum nanosensor parameters. The extinction spectra of the nanodisk arrays are measured using a homemade transmission microscopy set-up integrated with a VIS-NIR light source coupled to a spectrometer[29]. Our optical detection enables us to interrogate up to 32 regions in parallel throughout the chip for real-time resonance tracking of different sensor arrays.

The PDMS chip is fabricated by multilayer soft lithography leading to 8 sensing channels that are all individually and simultaneously addressable (see the Supplementary Information).[29] The sample flow on the experiment channels is controlled by the micromechanical valves[30] that are thin PDMS membranes between the flow layer and the control layer of the chip. By pressurizing the control channels, the thin PDMS membrane valve is actuated to control the flow of the samples on the experiment layer. These micromechanical valves are pressure-driven by electronic valves that are controlled by a home-made graphical user interface.

The detection of the biomolecules is based on a standard sandwich assay scheme where the protein of interest is captured between two specific antibodies, one immobilized on the silicon surface, acting as capture antibody and another that is flowed in solution as detection antibody (Fig. 1d). The capture antibody is immobilized on the silicon sensors by passive adsorption similarly to ELISA and other immunoassay techniques.[31] The samples, which contain different concentrations of the target protein of interest, are then introduced into individual channels and the protein is captured by the immobilized antibody. Finally, the detection antibody is introduced and is bound to the previously captured protein. This amplification step provides higher resonance shifts compared to the shifts obtained by the target proteins, offering higher concentration resolution for small proteins. It also increases the selectivity of the sensing protocol since the protein of interest is recognized by two specific antibodies (see the Supplementary Information). The increased effective refractive index at the sensor vicinity resulting from the binding steps (Fig. 1d) causes the resonances of the nanodisk arrays to redshift enabling accurate monitoring of the adsorption and binding kinetics of the molecules (Fig. 1e).

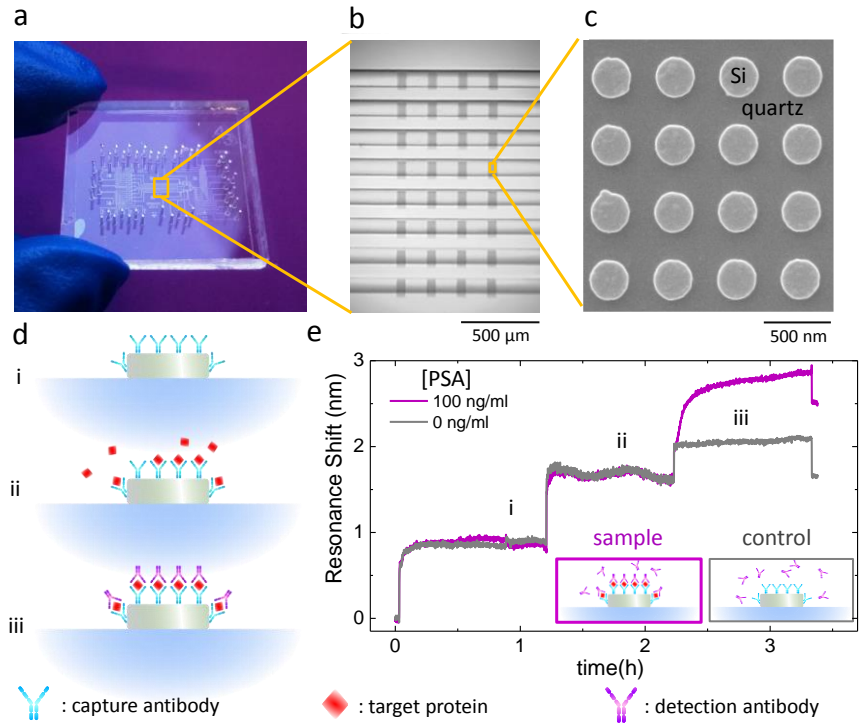

**Figure 1:** On-a-chip biosensing with silicon on quartz nanodisks. (a) Picture of an assembled chip with 8 sensing channels. (b) Close-up picture of the 8 sensing channels showing silicon nanodisk arrays with different parameters. (c) SEM micrograph of a small portion of a silicon nanodisk array with $h$=50nm, $r$=140 nm and $s$=200 nm. (d) Schematics of the sensing protocol. (e) Evolution of the nanodisk resonance during the different steps of the sandwich assay. The sample (100 ng/ml PSA) and control (no PSA) experiments are in grey and purple, respectively.

The extinction spectrum of individual $h$=50 nm Si nanodisks is mainly dominated by an electric dipole resonance (See Supplementary Fig. 1).[28,32] Here, for sensing purposes, we aim at exploiting the strong collective resonance arising from far field dipole coupling. The array resonance is optimized by changing the disk radius (*r*) and separation (*s*). Fig. 2 shows the measured extinction spectra of the different arrays along with the corresponding Finite Element Method (FEM) simulations. The resonances are tuned by changing the periodicity at fixed radius (Fig. 2a), or conversely, changing the nanodisk radius at constant disk separation (Fig. 2c). The corresponding FEM simulations on infinitely large arrays are in good agreement with the measured data (Fig 2b and 2d). The amplitude and width of each of these resonances vary for different nanodisk arrays. While these properties are important for the detectability of spectral shifts, also the refractive index sensitivity is expected to vary with the array parameters.

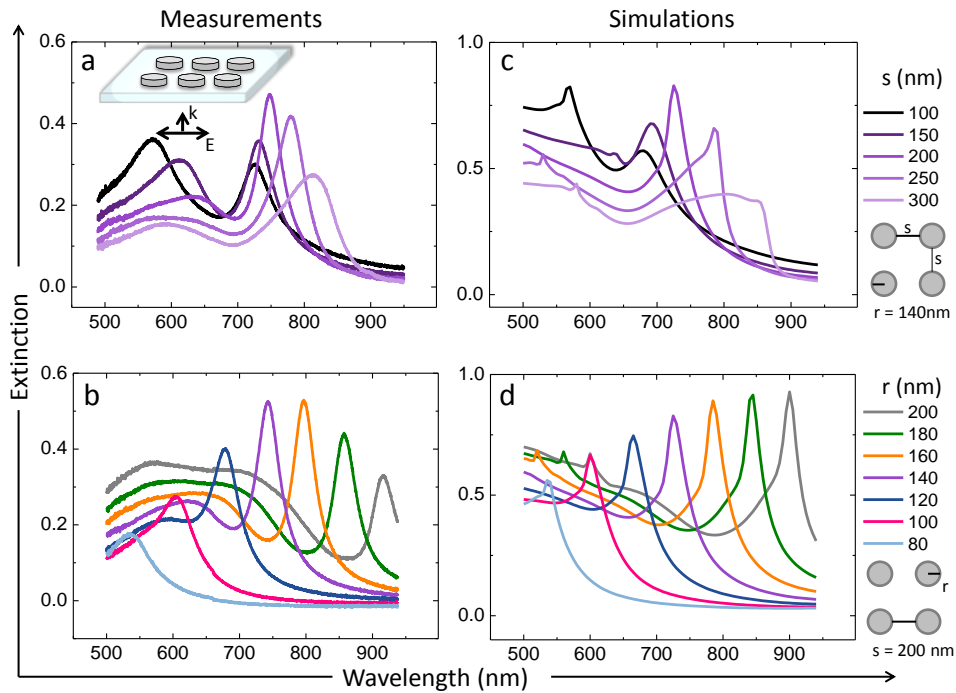

**Figure 2:** Resonance tuning of Si-nanodisk arrays. Experimental extinction spectra of silicon nanodisk arrays in air: (a) Influence of the disk separations at fixed radius *r*=140 nm and (b) Influence of the disk radius at fixed disk separation of *s*=200 nm. (c, d) Corresponding FEM simulations. Inset in (a) shows the geometry of the disks and the incident light polarization.

In order to identify the structural parameters (*r* and *s*) that provide the highest sensitivity to the surrounding media we performed bulk refractive index sensitivity (BRIS) experiments. In these experiments, we fabricated sensor arrays with 4 different radii ($r$=120, 140, 160 and 180 nm) and with disk separations varying from 100 nm to 450 nm with 50 nm increments. Once integrated to the microfluidics, the fabricated sensor arrays are exposed to increasing concentrations of glucose solutions in ultra-pure water (Fig. 3a).

For illustration, Fig. 3b shows the resulting redshift in the extinction spectra for $r$=140 nm and $s$=300 nm. Our automated parallel acquisition enables us to track in real time the spectrum of each of the different arrays on the chip and extract the corresponding shifts in the main peak centroid. For asymmetric extinction peaks like the one considered here, peak centroid tracking was found to be more sensitive than standard peak tracking (See Supplementary Fig. 2).[33] The inset of Fig. 3c displays the real time measurement of the centroid shifts for the same array where different glucose concentrations are flowed in the channels sequentially with a step of washing with water in between. We observe instantaneous shifts as the refractive index of the surrounding media changes. The redshifted signal returns back to the baseline value for the washing steps with water ensuring that there is no irreversible modification of the sensors and the shifts are indeed due to bulk refractive index changes.

Fig. 3c shows the evolution of the solution of the peak centroid with the refractive index for the same nanodisk array. From the slope of the linear fit we extract the bulk refractive index sensitivity (BRIS) of the sensor. The BRIS values for sensors with different disk separation and radius are gathered in Fig. 3d (See also supplementary Fig. 3). The arrays with separations larger than 300 nm were not considered as they either exhibit very low extinction, due to a low particle density, or resonances that were out of the spectral range of our set-up. Within the considered parameter range, we found that the BRIS values increased with increasing nanodisk separations. The highest BRIS value reached in our parameter range was 227 nm/RIU which corresponds to the BRIS of the array with $r$=140 nm and $s$=300 nm (Fig.

3b-d). Despite the simplicity of our structure, this BRIS value is only slightly lower than the previously reported BRIS values of more complex silicon nanostructure arrangements[22,23].

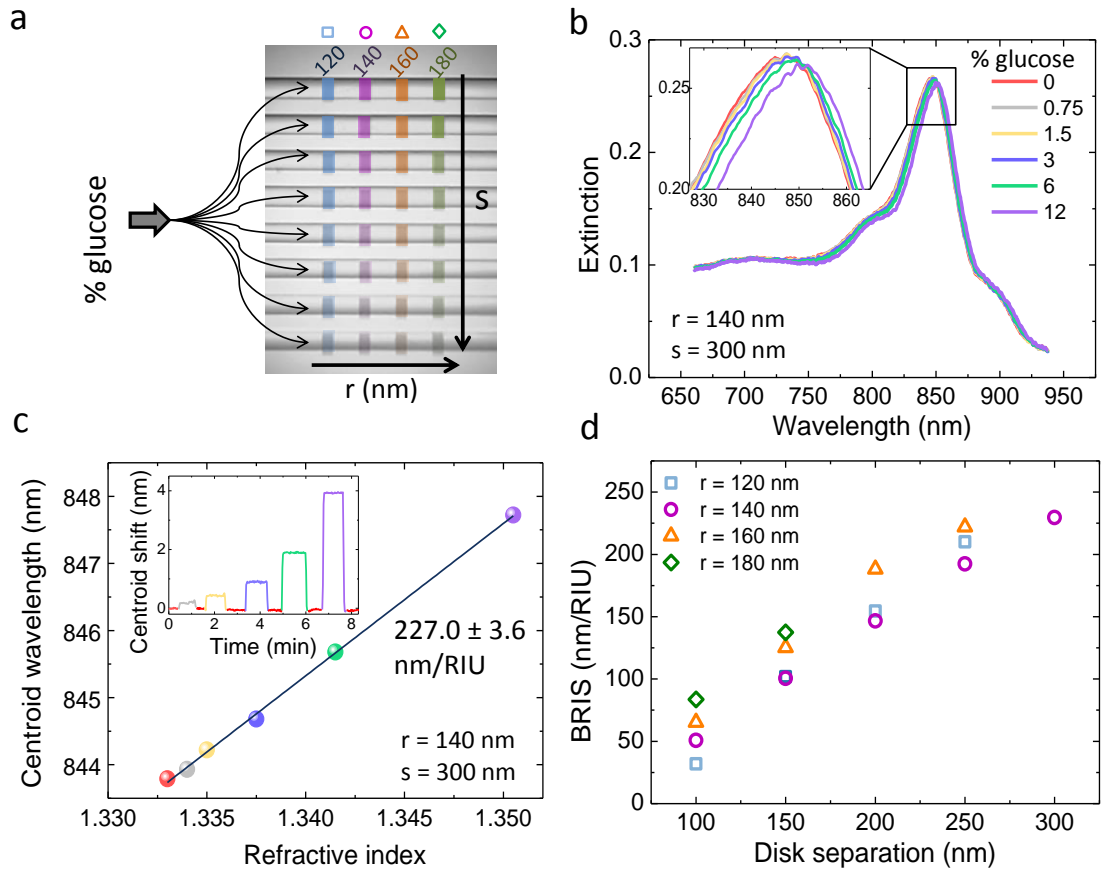

**Figure 3:** Bulk refractive index sensitivity (BRIS). (a) Schematics of the BRIS experiment in which the 8 microfluidic channels are used to flow different water/glucose mixtures on Si nanodisk arrays with different *s* and *r*. (b) Evolution of the extinction spectra of a nanodisk array (*r*=140 nm, *s*=300 nm) exposed to 6 different glucose-water mixtures. (c) The centroid positions extracted from the extinction spectra in b as a function of the refractive index of the glucose solutions. The real time centroid shifts during the sequential flow of varying glucose concentrations separated by rinsing step. (d) Summary of the BRIS values obtained for different arrays with different radii and disk separations. The error bars are smaller than the data points in the plot.

For the target analyte sensing proof of concept experiments we selected the two sensors exhibiting the highest BRIS values (with radii of 140 and 160 nm and disk separations of 300 and 250 nm, respectively). The experiment consists of flowing different concentrations of the target molecule in each of the individual channels in order to obtain an 8-point calibration curve. We here focused on the detection of Prostate Specific Antigen (PSA). PSA is a 34 kDa protein whose high concentration in blood (greater than 4-10 ng/ml) can be associated to prostate cancer or other prostate disorders.[34,35]

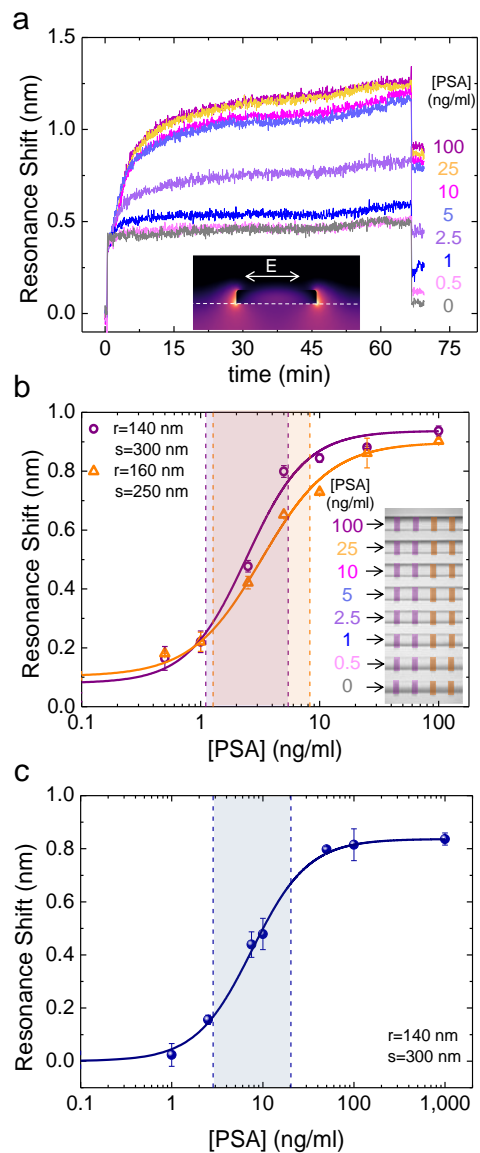

**Figure 4:** PSA detection (a) Real time resonance shifts of the silicon nanodisk arrays ($r$=140nm, $s$=300nm) due to the detection of different concentration of PSA. The inset shows

the calculated near field distribution for one nanodisk from an infinite array. The white dashed line indicates the substrate-nanodisk interface. (b) Calibration curves for PSA in PBS buffer with 1% BSA obtained using two different sensor arrays on the same chip (inset is a diagram showing the sensor organization in the channels). The shaded areas on the calibration curves represent the dynamic ranges of the curves (purple: $r$=140 nm, $s$=300 nm array, orange: $r$=160 nm, $s$=250 nm array). The error bars indicate the standard deviation of the 2 parallel measurements on the same chip. (c) PSA calibration curve obtained in 50% diluted human serum for sensor with $r$=140 nm and $s$=300 nm. The shaded area defines the dynamic range and the error bars are the standard deviations of the 4 different measurements on the same chip.

For the detection of PSA, the capture antibody is first immobilized on the sensor surface by passive adsorption by flowing the antibody solution in phosphate buffer through all 8 channels. The binding kinetics of the antibody to the sensor surface is monitored by tracking the peak centroid shifts of the silicon nanodisks. Once the signal saturation is observed, the sensors are washed with PBS buffer. Then, different concentrations of PSA dissolved in the assay buffer (PBS with 1 % BSA) are flowed in each individual channels. After another washing step with PBS, the detection antibody for PSA is introduced on the sensor surface for the detection step.

Fig. 4a shows the real time recorded shifts of the sensors with radius of 140 nm and separation of 300 nm in the detection step. The channels are washed with PBS to remove unbound detection antibodies and reduce the bulk refractive index effect. This washing step is also seen in Fig. 4a, as a drop of signal after the detection antibody binding curves have reached the plateau. The final shifts after the washing step are extracted to obtain the calibration curves for PSA (Fig. 4b). The standard curves were fitted using a four-parameter logistic equation. The analytical parameters of the sandwich assay are shown in Supplementary Table 1.

The dynamic ranges of the sensors are shown by the shaded areas on the calibration curves which cover the clinically relevant range for both of the sensors. The limit of detection (LOD, estimated as the conventional $IC_{10}$ value of the four-parameter logistic curve fit) for both sensors are beyond the cut-off concentration for patients. The LOD for $r$=140 nm and $r$=160 nm sensors are around 0.69 ng/ml and 0.74 ng/ml, respectively. The sensitivity of a sensor is conventionally defined as the $IC_{50}$ value of the calibration curve which is found to be 2.45 ng/ml and 3.24 ng/ml, respectively. We find that the sensors with higher BRIS ($r$=140 nm, $s$=300 nm) led to calibration curves with slightly better sensitivity and lower LOD values with a higher slope of the linear range of the sensors (Supplementary Table 1). The error bars are representing the signal variations between the replicas on the same chip and the relative standard deviation for intrachip reproducibility is found to be varying between 0.5% and 5.1% for the working range of the sensors.

Finally, to demonstrate the clinical relevance of our sensing platform, we performed the PSA calibration curve measurements in human serum (Fig. 4c). Serum was diluted in at 50% in PBS in order to reduce the matrix effects. The LOD ($IC_{10}$) is 1.6 ng/ml which is below the cut-off value of the PSA concentration in patients and the dynamic range is between 2.5 ng/ml and 16.0 ng/ml which covers the clinically relevant range for cancer screening. Further details about the analysis of the calibration curve can be found in Supplementary Table 1. This result suggests that the sensor performance is within the clinically relevant range and the measurements are feasible even in a complex matrix such as human serum. Very low shifts of the control channel (0.005 ± 0.035 nm) indicates negligible unspecific signal.

In order to compare the sensing performance of our silicon nanoresonators with their plasmonic counterparts[11–13], we repeated the PSA calibration curve experiment with the optimized gold nanorod array studied in our previous work[29]. The only significant difference comes from the use of EDC/NHS based surface chemistry prior to antibody immobilization on gold. Otherwise, both measurements were performed under the same conditions, such as antibody concentrations, buffers and similar flow times.

The normalized PSA calibration curves for the gold nanorods (100 nm x 200 nm) and the silicon nanodisks ($r$=140 nm, $s$=300 nm) are compared in Fig. 5. The analytical parameters of the sandwich assay on both sensors can be found in Supplementary Table 1. The LOD (IC$_{10}$) of the gold sensors is found to be around 0.87 ng/ml which is slightly higher than the LOD of the silicon based sensors. Conversely, the dynamic range of the LSPR-based sensors is broader for higher concentrations, giving a lower slope, which leads to higher limit of quantification. Besides, the assay time was much shorter for silicon based sensors since the LSPR sensing protocol requires self-assembled monolayer preparation and activation as well as blocking steps using ethanolamine (Supplementary Fig. 4).

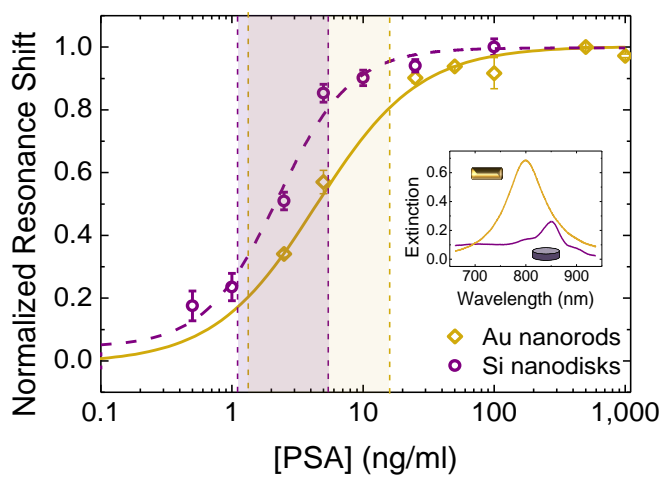

**Figure 5:** Comparison of the Si and Au platforms. PSA calibration curves for gold nanorod arrays (gold) and silicon nanodisk arrays (purple). Both curves were normalized for clearer comparison. The error bars represent the standard deviation of 2 parallel measurements on the same chip. The shaded areas are the respective dynamic ranges. The inset shows the experimental extinction spectra of gold nanorod arrays and silicon nanodisk arrays.

We have studied the use of silicon nanoresonators integrated in a state-of-the-art microfluidic architecture for biosensing. We demonstrated that the resulting platform is compatible with detection of small biomolecules in complex matrices for clinical applications. The reported sensing performance enables us to detect clinically relevant concentrations of PSA.

We have also compared our platform with a well-developed LSPR-based sensing protocol and shown that the sensitivity, LOD and the dynamic ranges are comparable. Besides the similar sensing performance, one of the advantages of the silicon based sensors is the significantly longer decay length of the surface field over LSPR modes in metal nanoantennas (Supplementary Fig. 1), which can be beneficial for multilayer assays involving detection of molecules relatively far from the surface. This suggests that in comparison with LSPR sensors, different assay types with multiple layers of antibodies can be efficiently monitored using the silicon nanoresonators which in practice enables more practical and faster detection of the target analyte.

The extinction of the Si nanoresonators considered here is much lower compared to the extinction of the gold nanoantennas (inset Fig. 5). This suggests that for Si resonators there is still room for improvement by tailoring the shape, size, periodicity and height of nanoresonator arrays in order to obtain sharper, stronger and potentially more sensitive resonances. We thus foresee further engineering of the Si nanoresonators including arrangement in dimers and oligomers, that could improve the sensing performance. However, more complex nanostructuring of the silicon might also affect the local sensitivity to molecular adsorption at the surface of the nanoresonators. While the location of the adsorbed antibodies is not controlled in the present study (see Supplementary Fig. 5 for a simulation of the comparison of various cases), this parameter can become important when studying more complicated designs.

Finally, the quality of the sandwich assay, hence the LOD and sensitivity, can further be improved by additional optimizations of the capture and detection antibody concentrations.

## ASSOCIATED CONTENT

Supporting Information

Sensor fabrication, Microfluidic chip fabrication, Measurements, Sandwich assay formation on chip, Simulations, Supplementary Figures, Supplementary Table.

## AUTHOR INFORMATION

**Corresponding Author**

Romain Quidant, romain.quidant@icfo.es

**Author Contributions**

R.Q. proposed the idea and co-supervised the work with V.S. O.Y. designed the silicon sensor arrays and fabricated the on-chip devices, carried out the measurements and processed the data. M.S. carried out the FEM simulations and interpreted the resonance modes. P.D. and V.S. supervised the sandwich assay chemistry. All authors analyzed the data and contributed to the writing of the manuscript.

**Notes**

The authors declare no competing financial interests.

## ACKNOWLEDGEMENTS

The authors acknowledge financial support from the European Community's Seventh Framework Program under grant QnanoMECA (64790), Fundació Privada Cellex, and the Spanish Ministry of Economy and Competitiveness, through the 'Severo Ochoa' Programme for Centres of Excellence in R&D (SEV-2015-0522) and grant FIS2013-46141-P and Swedish Research Council (637-2014-6894).